\documentclass[conference]{IEEEtran}
\IEEEoverridecommandlockouts

\usepackage{cite}
\usepackage{amsmath,amssymb,amsfonts}
\usepackage{algorithmic}
\usepackage{graphicx}
\usepackage{textcomp}
\usepackage{xcolor}
\usepackage{multirow}
\usepackage{threeparttable}
\def\BibTeX{{\rm B\kern-.05em{\sc i\kern-.025em b}\kern-.08em
    T\kern-.1667em\lower.7ex\hbox{E}\kern-.125emX}}
\begin{document}

\title{Path-SAM2: Transfer SAM2 for digital pathology semantic segmentation\\
}

\author{
Mingya Zhang$^{1} $  \quad Liang Wang$^{1} $ \quad Zhihao Chen$^{2}$ \quad Yiyuan Ge$^{2}$ \quad Xianping Tao$^{1}$\\
    $^{1}$ Nanjing University \quad
    $^{2}$ BISTU \quad
}

\maketitle

\begin{abstract}
The semantic segmentation task in pathology plays an indispensable role in assisting physicians in determining the condition of tissue lesions.
With the proposal of Segment Anything Model (SAM), more and more foundation models have seen rapid development in the field of image segmentation. 
Recently, SAM2 has garnered widespread attention in both natural image and medical image segmentation. Compared to SAM, it has significantly improved in terms of segmentation accuracy and generalization performance.
We compared the foundational models based on SAM and found that their performance in semantic segmentation of pathological images was hardly satisfactory.
In this paper, we propose Path-SAM2, which for the first time adapts the SAM2 model to cater to the task of pathological semantic segmentation. We integrate the largest pretrained vision encoder for histopathology (UNI) with the original SAM2 encoder, adding more pathology-based prior knowledge. Additionally, we introduce a learnable Kolmogorov–Arnold Networks (KAN) classification module to replace the manual prompt process.
On three adenoma pathological datasets, Path-SAM2 has achieved state-of-the-art performance.
This study demonstrates the great potential of adapting SAM2 to pathology image segmentation tasks.
We plan to release the code and model weights for this paper at: https://github.com/simzhangbest/SAM2PATH
\end{abstract}


\section{Introduction}

\begin {figure*}[ht]
\centering
\includegraphics[width=\textwidth]{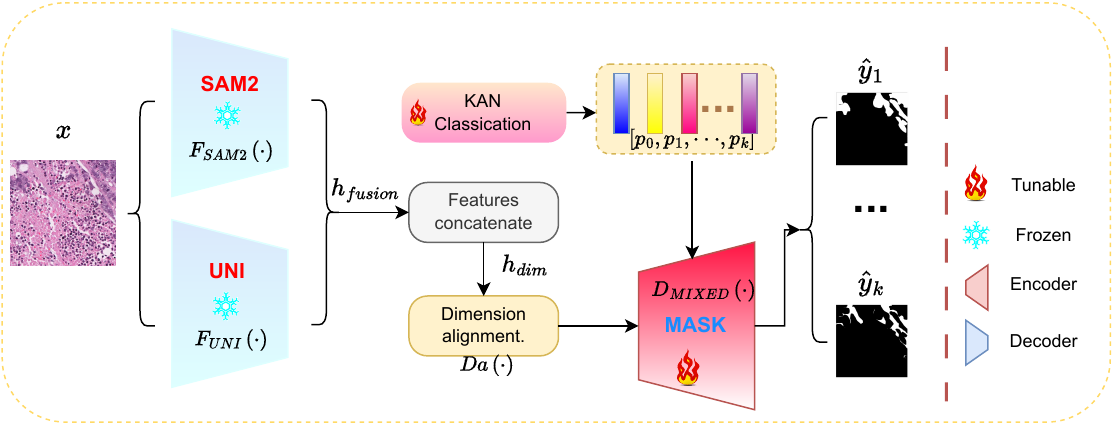}
\centering

\caption{Overview of Path-SAM2. The concatenated features from SAM2 and UNI are then passed to a Dimension Alignment module $Da\left ( \cdot  \right ) $. K learnalbe prompt tokens are produced from KAN Classification module.}
\label{fig:sam-uni}
\end {figure*}

Digital pathology has transformed the field of histopathological analysis by employing advanced computational methods to enhance the diagnosis and prognosis of diseases~\cite{Litjens2017ASo,Tizhoosh2018ArtificialIA}.
Semantic segmentation is a pivotal technique in digital pathology, which involves segmenting images into distinct regions that represent various tissue structures, cell types, or subcellular elements~\cite{Komura2018MachineLM,Zhang2023PreciseLM,Kapse2022SubtypeSpecificSD}.
The precision and efficiency of semantic segmentation are crucial for a wide range of applications, including the detection, grading, and prediction of outcomes for tumors, as well as for the examination of tissue morphology and cell interactions. Consequently, the development and refinement of robust segmentation algorithms are of paramount importance for the continued progress in the field of digital pathology~\cite{Niazi2019DigitalPA,Ding2022ImageAR,Lu2021FeaturedrivenLC}.

Large foundation  models have shown great promise in a multitude of fields, thanks to their powerful zero-shot learning abilities~\cite{wei2022emergent, kirillov2023segment, wang2023seggpt, wang2023images}. Particularly in the domain of medical image segmentation, the Segment Anything Model (SAM)~\cite{kirillov2023segment} has made impressive strides in zero-shot segmentation tasks, known as MedSAM~\cite{MedSAM}. 
SAM2~\cite{ravi2024sam2} is an upgraded version of SAM, with added video segmentation capabilities and improved segmentation accuracy. MedSAM2~\cite{ma2024segment} has also enhanced the accuracy of zero-shot segmentation for medical images and added segmentation for medical videos.
UNI~\cite{chen2024uni} is a cutting-edge self-supervised model for computational pathology, pre-trained on over 100 million images from a diverse range of diagnostic $H\&E$-stained whole-slide images. 
SAM-Path~\cite{zhang2023sam} incorporates the encoder of HIPT~\cite{Chen2022ScalingVT} model as an external pathology encoder into the workflow of SAM.

Incorporating the above information, we believe that in the field of pathological image segmentation, using an external encoder that contains more prior pathological knowledge can better utilize SAM2's image processing capabilities.

Recently, the advent of Kolmogorov-Arnold Networks (KANs) has aimed to demystify the opaque nature of traditional neural network designs, offering enhanced interpretability and showcasing the promise of transparent AI research~\cite{yu2024white,pai2024masked}. 
Leveraging the flexibility and high accuracy of the KAN architecture, we utilize the KAN classification prompt module in Path-SAM2, thereby endowing our network architecture with more precise classification capabilities.


Our primary contributions are summarized as follows:
1. We pioneer the development of a pathology image segmentation model known as Path-SAM2, which is the first of its kind to utilize a SAM2 foundation.
2. By incorporating the largest existing pathology model, UNI, as an extra encoder, we enhance SAM2's ability to assimilate domain-specific knowledge in pathology.
3. We introduce an innovative trainable prompt method, inspired by KAN, which empowers SAM2 to execute semantic segmentation tasks with precision.

\section{Method}
\label{sec:method}

As shown in figure~\ref{fig:sam-uni}, our method is composed of five modules: a SAM2 image encoder $F_{SAM2}\left ( \cdot \right ) $, a pathology image encoder $F_{UNI}\left ( \cdot \right ) $, a dimensional alignment module $Da\left ( \cdot  \right ) $, a KAN classification module, and a hybrid decoding module $D_{MIXED}\left ( \cdot \right ) $.
For a given pathology image $x$, our prediction task is to obtain its corresponding semantic segmentation mask map $\hat{y}$, which has the same resolution as $x$. Each pixel in $\hat{y}$ corresponds to category information. Therefore, the predicted output sequence $\left [ \hat{y}_{1},\hat{y}_{2}, \cdot\cdot\cdot,\hat{y}_{i},\cdot\cdot\cdot,\hat{y}_{k} \right ] $, where $i$ represents the category information, and $k$ is the total number of categories.

\subsection{Pathology encoder}
SAM2 utilizes the Hiera~\cite{ryali2023hiera} network (a hierarchical vision transformer) to process natural images in daily life, and its encoder part indeed possesses the ability to understand pathology. In this paper, we employ an additional pathological encoding module to supplement SAM2 with more knowledge within the field of pathology.

The pathological encoder module we use originates from UNI~\cite{chen2024uni}, which is a versatile self-supervised model designed for pathology. It is pre-trained on over 100 million images derived from more than 100,000 diagnostic $H\&E$ stained whole-slide images (WSIs), covering 20 major tissue types and encompassing over 77 terabytes of data. UNI has been evaluated across 34 diverse computational pathology tasks with varying levels of diagnostic complexity.

The input pathological images are first processed by the SAM2 encoder and the UNI encoder, and their output features are concatenated in the dimensionality. The expression is as follows: 

\begin{equation}
    h_{fusion} = concat\left ( F_{SAM2}\left ( x \right ), F_{UNI}\left ( x \right )   \right ) 
\end{equation}
The encoder of SMA2 includes a neck section for dimension reduction, which is used to adjust the dimensions for isometric computation with the decoder. Because we have employed an additional encoder with pathological knowledge, the Path-SAM2 has removed the neck network from SAM2, and after concatenating the output features of the encoders, the network inputs the concatenated features into the Dimension Alignment(DA) module.

\subsection{KAN classification prompts}

\begin{figure}[hbpt]
    \centering
    \includegraphics[width=\linewidth]{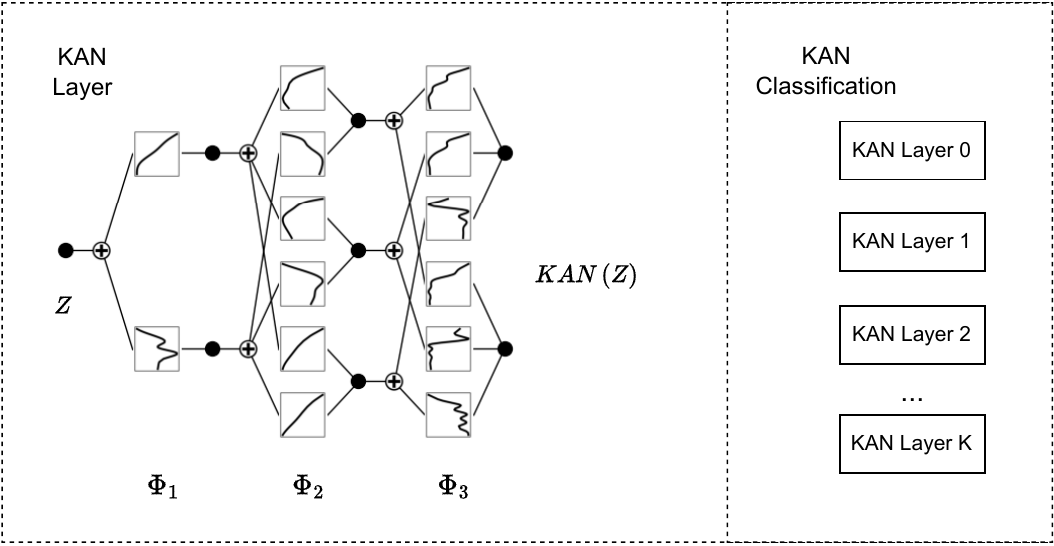}
    \caption{Overview of KAN Classification: This module is composed of k KAN Layers, each KAN Layer replaces a traditional MLP (Multi-Layer Perceptron) module, providing a trainable prompt for the segmented types.}
    \label{fig:kan}
\end{figure}

The Kolmogorov-Arnold (K-A) theorem~\cite{kolmogorov1957representation} suggests that any continuous function can be constructed from a combination of continuous single-variable functions, offering a foundation for universal neural network models. Hornik et al. ~\cite{hornik1989multilayer} confirmed this by showing that feed-forward networks can approximate any continuous function, a key principle in deep learning's evolution. 



To address the challenges of inefficient parameter utilization and poor interpretability typically found in Multi-Layer Perceptrons (MLPs), Liu and colleagues introduced the Kolmogorov-Arnold Network (KAN)~\cite{huang2014deep}. This new architecture is informed by the Kolmogorov-Arnold representation theorem and, like an MLP, it features a KAN with K layers that are structured as a series of nested KAN sub-layers:
\begin{equation}
    \text{KAN}(Z) = \left(  \Phi_{K-1}\circ \Phi_{K-2}\circ\cdot\cdot\cdot\Phi_{1}\circ\Phi_{0} \right) Z,
\end{equation}
In the context of the KAN network, $\Phi_{i}$ denotes the $i$-th stratum within the complete architecture. A single KAN stratum, which accepts an input of dimension $n_{in}$ and produces an output of dimension $n_{out}$, is constituted by a matrix of $n_{in} \times n_{out}$ trainable activation functions, represented by $\phi$.
\begin{equation}
    \Phi=\left \{ \phi_{q,p} \right \} , p=1,2,\cdot\cdot\cdot,n_{in}, q= 1,2,3,\cdot\cdot\cdot,n_{out}
\end{equation}
The computation result of the KAN network from layer $k$ to layer $k+1$ can be expressed in matrix form:
\begin{equation}
    \resizebox{\linewidth}{!}{
$        \mathbf{Z}_{k+1}=\underbrace{\left(\begin{array}{cccc}
        \phi_{k, 1,1}(\cdot) & \phi_{k, 1,2}(\cdot) & \cdots & \phi_{k, 1, n_{k}}(\cdot) \\
        \phi_{k, 2,1}(\cdot) & \phi_{k, 2,2}(\cdot) & \cdots & \phi_{k, 2, n_{k}}(\cdot) \\
        \vdots & \vdots & & \vdots \\
        \phi_{k, n_{k+1}, 1}(\cdot) & \phi_{k, n_{k+1}, 2}(\cdot) & \cdots & \phi_{k, n_{k+1}, n_{k}}(\cdot)
        \end{array}\right)}_{\boldsymbol{\Phi}_{k}} \mathbf{Z}_{k},$
        }
\end{equation}
In summary, KANs set themselves apart from standard MLPs through the incorporation of edge-based learnable activation functions and the use of activation functions as parametric weights, thereby sidelining the requirement for conventional linear weight matrices. 

As shown in Figure 2, to eliminate the need for manual input of prompts in the decoder Dmixed, we utilize trainable prompt tokens instead of human prompts. In the semantic segmentation task with k categories, we train a set of prompts, and the tokens of these prompts can be represented as $P = \left [ p_{1},p_{2},\cdot\cdot\cdot,p_{i},\cdot\cdot\cdot,p_{k} \right ] $, 
where $p_{i} $ is the prompt token for the $i_{th}$ category. The classification prompt of SAM-PATH2 is composed of $k$ KAN Layers, each of which is a complete KAN network.

For a segmented type $p_{i}$, the decoder of Path-SAM2 generates a corresponding segmentation mask $y_{predi}$ and an IOU parameter $iou_{i}$ for the $i_{th}$ class. The prediction process of the decoder is as follows:
\begin{equation}
    D_{MIXED}\left ( h_{dim}^{'},P \right ) = \left \{ \left \langle iou_{i}, y_{predi} \right \rangle \mid  i = 1,\cdot\cdot\cdot,k \right \} 
\end{equation}

\subsection{Loss Function}
The SAM employs a hybrid loss function that integrates Dice loss, focal loss, and IOU loss (which is essentially an MSE loss applied to IOU predictions). We have modified their loss function in the following:
\begin{equation}
    \mathcal{L}=\sum_{i=1}^{k}\left[(1-\alpha) \mathcal{L}_{\text {dice }}+\alpha \mathcal{L}_{\text {focal }}+\beta \mathcal{L}_{\text {mse }}\right]
\end{equation}
where $\alpha$ and $\beta$ are weight hyper parameters. $\mathcal{L}_{\text {dice}}$ represents the Dice loss function, $\mathcal{L}_{\text {focal}}$ represents the focal loss function and $\mathcal{L}_{\text {mse}}$ represents the Mean Squared Error loss function. 


\section{EXPERIMENTS AND RESULTS}
\label{experiments}

\subsection{Dataset}

In this paper, we evaluate the model using three pathology slide datasets: the EBHI~\cite{shi2023ebhi}, CRAG~\cite{Graham2019MildNetMIL}, and GlaS~\cite{sirinukunwattana2017gland}.
For these datasets, we adhere to the official division between training and testing data, and additionally, we allocate $20\%$ of the training data to form a distinct validation set.

\noindent\textbf{EBHI:} The dataset encompasses 4,456 $H\&E$-stained images, which include six distinct categories of histological section images along with their respective accurate annotation images, all measuring $224 \times 224$ pixels.

\noindent\textbf{CRAG:} The Colorectal Adenocarcinoma Gland dataset features 213 images, each approximately $1536 \times 1536$ pixels in dimension, extracted from 38 whole slide images of $H\&E$-stained samples at a 20x magnification.

\noindent\textbf{GlaS:} The dataset is composed of 165 images, originating from 16 $H\&E$-stained histological sections of colorectal adenocarcinoma, with each image sized at $522 \times 775$ pixels.

\subsection{Results}

\begin{table}[]
\caption{Quantitative results of segmentation on the three datasets }
\label{tab:results}
\begin{threeparttable}
\resizebox{\columnwidth}{!}{%
\begin{tabular}{|l|ll|ll|ll|}
\hline
\multicolumn{1}{|c|}{\multirow{2}{*}{Models}} & \multicolumn{2}{c|}{EBHI} & \multicolumn{2}{c|}{CRAG} & \multicolumn{2}{c|}{GlaS} \\ \cline{2-7} 
\multicolumn{1}{|c|}{} & DSC & IOU & DSC & IOU & DSC & IOU \\ \hline
Vanilla SAM & 50.17 & 38.33 & 52.45 & 35.55 & 49.73 & 37.11 \\ \hline
Vanilla SAM ( pp\tnote{1} ) & 70.23 & 62.51 & 65.95 & 49.24 & 57.28 & 43.29 \\ \hline
Vanilla SAM2 & 55.64 & 45.29 & 56.83 & 40.81 & 53.79 & 42.43 \\ \hline
Vanilla SAM2( pp\tnote{1} ) & 77.95 & 69.25 & 80.98 & 69.76 & 79.05 & 67.68 \\ \hline
Fine-tuned SAM & 56.19 & 42.38 & 57.32 & 40.26 & 56.41 & 42.69 \\ \hline
Fine-tuned SAM2 & 58.87 & 50.24 & 62.43 & 53.17 & 59.29 & 47.82 \\ \hline
MedSAM2( pp\tnote{1} ) & 76.04 & 62.29 & 65.29 & 49.72 & 64.19 & 48.55 \\ \hline
SAM-Path & 90.04 & 91.26 & 88.41 & 88.31 & 84.71 & 89.92 \\ \hline
Path-SAM2(Ours) & \textbf{92.23} & \textbf{93.17} & 88.09 & \textbf{89.38} & \textbf{85.3} & \textbf{92.02} \\ \hline
\end{tabular}%
}
\begin{tablenotes}
    \footnotesize
    \item[1] "pp" stands for manual dot prompts and post-processing the mask obtained from the SAM models to support semantic segmentation.
\end{tablenotes}
\end{threeparttable}
\end{table}

\subsubsection{Implementation details}

\begin {figure}[hbpt]
\centering
\includegraphics[width=\linewidth]{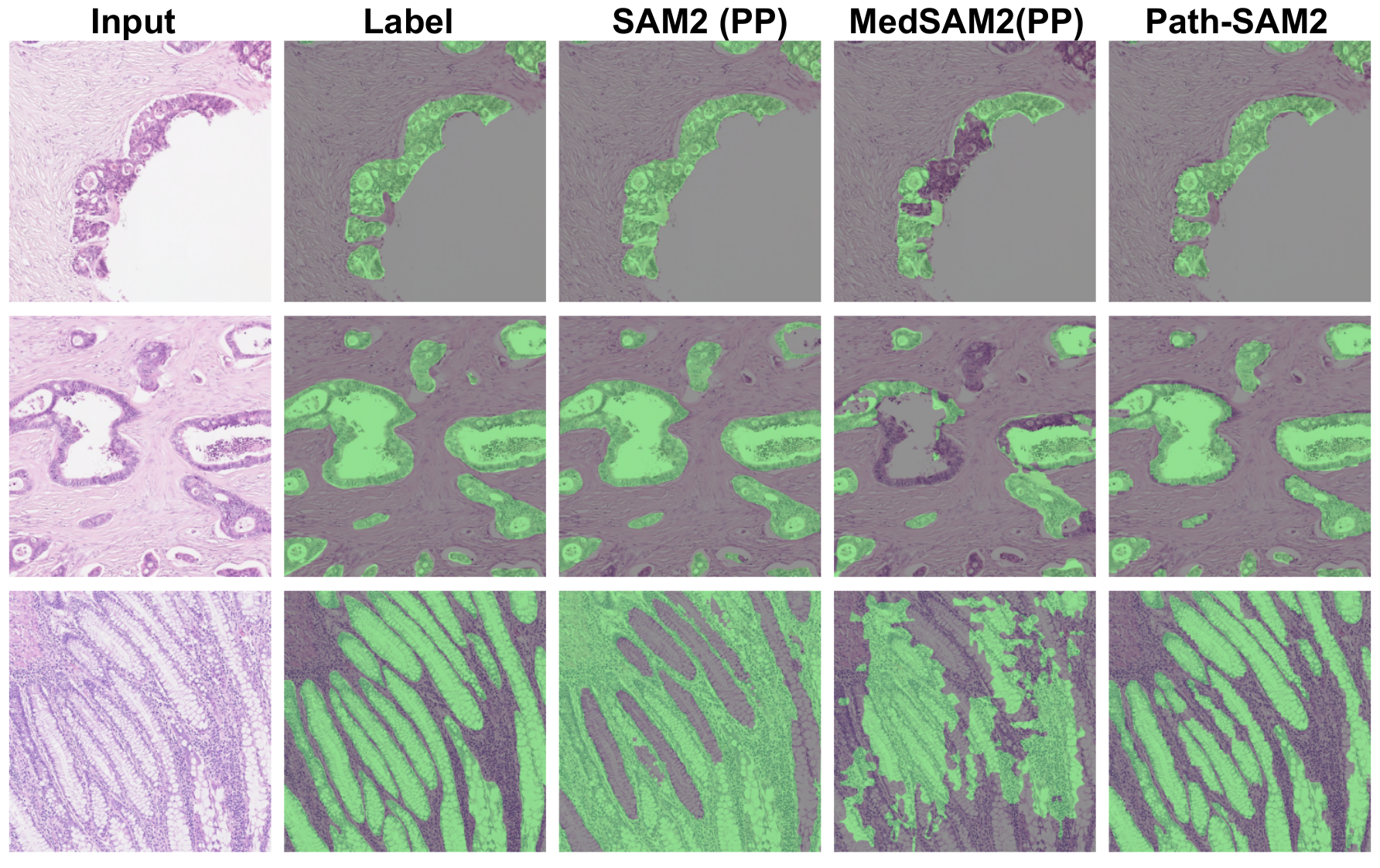}
\centering

\caption{Qualitative analysis on the CRAG dataset}
\label{fig:crag}
\end {figure}

\begin {figure}[hbpt]
\centering
\includegraphics[width=\linewidth]{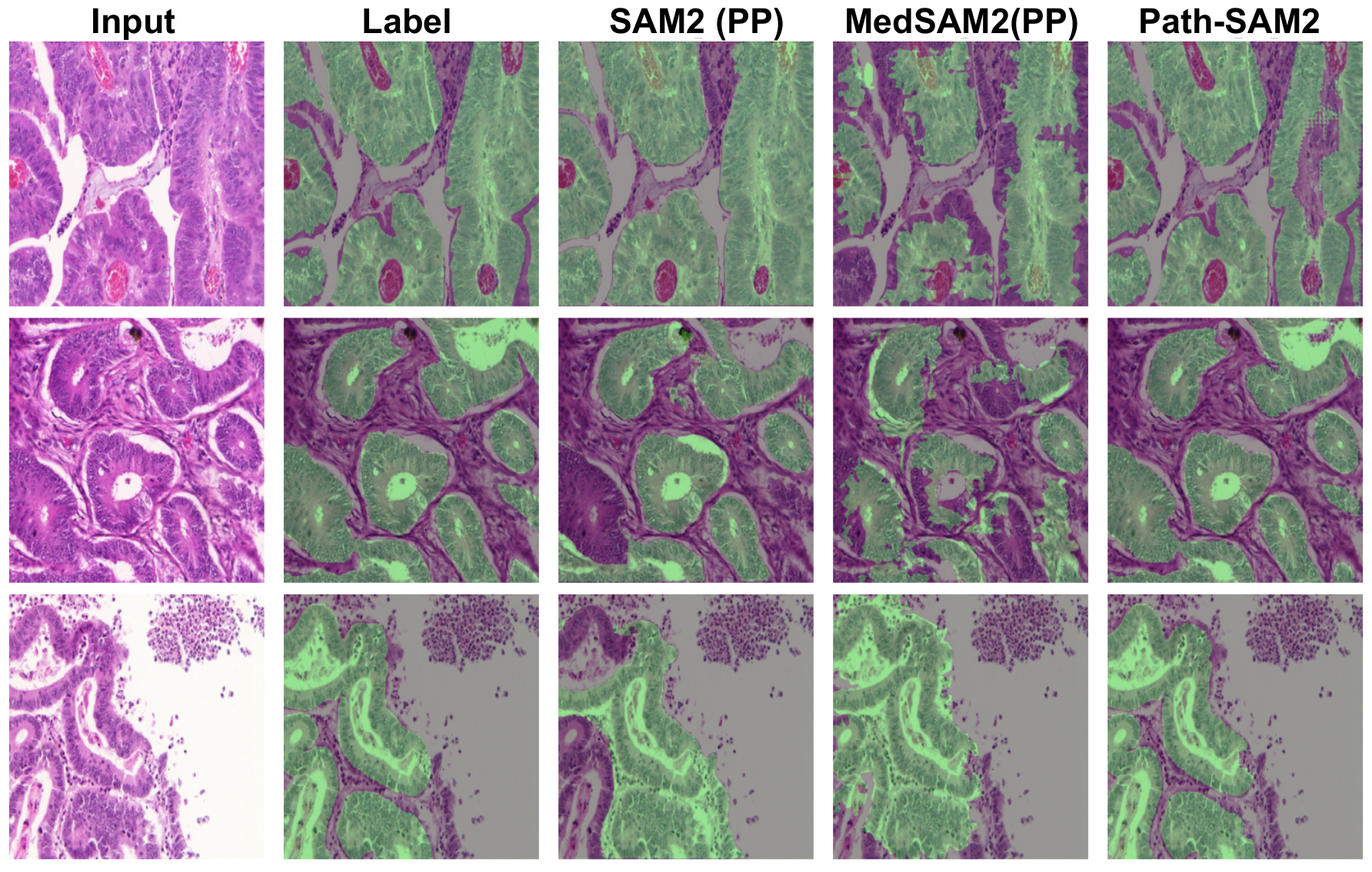}
\centering

\caption{Qualitative analysis on the GlaS dataset}
\label{fig:glas}
\end {figure}

\begin {figure}[hbpt]
\centering
\includegraphics[width=\linewidth]{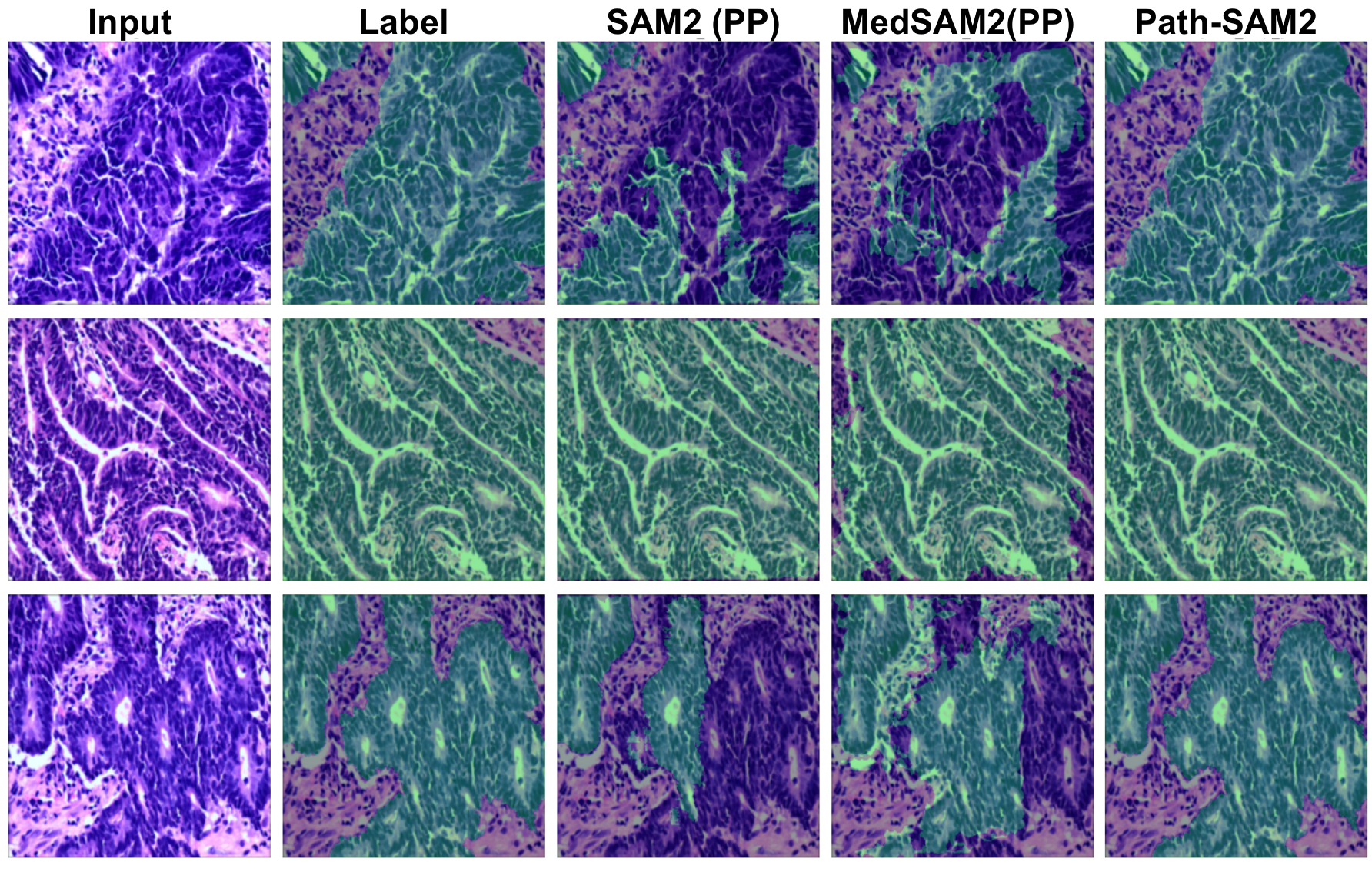}
\centering

\caption{Qualitative analysis on the EBHI dataset}
\label{fig:ebhi}
\end {figure}

We used AdamW~\cite{loshchilov2017adamW} optimizer with a weight decay of 1e-2 and 1e-5 learning rate.
The pre-trained models we used include SAM2 and UNI, and we conducted experiments using their base versions, employing a three layers KAN classification network.
For the hyperparameters of the loss function, we set  $\alpha$ = 0.125, $\beta$ = 0.01
We used the PyTorch library and trained our network on four Nvidia RTX V100 GPUs.

\subsubsection{Quantitative analysis}

For all datasets, we employ the Dice Similarity Coefficient(DSC) and Intersection Over Union (IOU) as our primary metrics for evaluation. 
We use SAM~\cite{kirillov2023segment}, SAM2~\cite{ravi2024sam2}, MedSAM2~\cite{ma2024segment}, and SAM-path~\cite{zhang2023sam} as our baselines, where vanilla SAM refers to the manual dot prompt SAM model, and SAM with post-processing refers to further processing the mask obtained from the vanilla SAM. The same operations are also performed on SAM2 and MedSAM2. 
The fine-tuning of SAM and SAM2 involves not using the UNI encoder, but rather incorporating our KAN classification module into the original SAM2 encoder and decoder for training.
When calculating metrics using the baseline, the differences in manual dot prompts can greatly affect the results. We choose the best prompt within our capabilities, which is to involve pathologists to assist in the manual dot prompt.

As shown in Table~\ref{tab:results}, "SAM with post-process" and the fine-tuned SAM enhance the segmentation performance of the vanilla SAM. SAM2, when subjected to the same operations, will also improve segmentation performance. MedSAM2, which has been fine-tuned with a large number of medical images and masks, is expected to yield better results in the segmentation of pathological images. However, their performance is not very satisfactory, and they require a significant amount of manual prompt and post-processing work, with a high degree of human involvement and unstable performance.

Compared to various processing methods for SAM, Path-SAM initially proposed enhancing the encoding capability of the SAM encoder by adding an encoder structure. Inspired by this approach, in Path-SAM2, we have, for the first time, tailored and integrated the encoder part of the UNI model into the SAM2 network. This significantly enhances the generalization ability of SAM2 in the field of pathological image analysis. Data indicates that on three public pathological datasets, our Intersection over Union (IOU) metric is higher than the baselines.
Taking the results on the EBHI dataset as an example: compared to vanilla SAM2 and SAM2 with post-processing, the IOU has improved by 47.88$\%$ and 23.92$\%$, respectively.

\subsubsection{Qualitative analysis}
We have partially visualized the segmentation results on the three datasets for comparison with the baselines. As shown in Figure~\ref{fig:crag}, Figure~\ref{fig:glas}, and Figure~\ref{fig:ebhi}, we consider adenomas in pathological images as the type to be segmented. Due to the lack of pathology-related prior knowledge in SAM and SAM2, we compared the post-processing parts of vanilla SAM and vanilla SAM2 and found that, despite the carefully designed manual dot prompts, the model still forcibly omitted parts of the adenoma area or misidentified non-adenoma parts as adenomas.
After fine-tuning, MedSAM2, with post-processing, significantly improved the previously mentioned issues, and the segmentation results approached satisfaction. From our Path-SAM2 segmentation images, it can be seen that the addition of an extra pathology-related UNI encoder has brought the semantic segmentation effect very close to the ground truth label. This also proves that the introduction of an additional encoder can significantly enhance the segmentation performance of SAM2 in downstream tasks.

\subsubsection{Ablation study}
Regarding the overall architecture of Path-SAM2, we have already discussed the comparison of the encoder part in the previous section. In this section, we discuss the impact of the KAN classification module on the model's performance, and we will compare it with the MLP (Multilayer Perceptron).
As shown in Table~\ref{tab:kan}, where "w.o" is the abbreviation for "without" and "w." is the abbreviation for "with". when using the KAN classification module, we achieved superior IOU and Dice scores on three public pathology datasets compared to not using this module.

\begin{table}[]
\caption{The ablation study of Path-SAM2}
\label{tab:kan}
\resizebox{\columnwidth}{!}{%
\begin{tabular}{|l|llllll|}
\hline
\multirow{2}{*}{} & \multicolumn{2}{l}{EBHI} & \multicolumn{2}{l}{CRAG} & \multicolumn{2}{l|}{GlaS} \\ \cline{2-7} 
 & DSC & IOU & DSC & IOU & DSC & IOU \\ \hline
w.o. KAN & 92.17 & 92.23 & 83.03 & 86.29 & 71.45 & 89.8 \\ \hline
w. KAN & \textbf{92.23} & \textbf{93.17} & \textbf{88.09} & \textbf{89.38} & \textbf{85.3} & \textbf{92.02} \\ \hline
\end{tabular}%
}
\end{table}

\section{conclusion}
In this paper, we propose Path-SAM2, which introduces the largest pre-trained model in the pathology field UNI, and adds the KAN classification module to replace manual dot prompts, achieving pathology semantic segmentation based on SAM2. We have validated it on three public pathology datasets, and Path-SAM2 has achieved the best segmentation results in terms of DSC (Dice Similarity Coefficient) and IOU (Intersection over Union) metrics compared to the baseline models. Our work confirms the potential of SAM2 in the research of semantic segmentation of pathological images.

\bibliographystyle{IEEEbib} 
\bibliography{refs}

\end{document}